\documentclass[prb,twocolumn,groupeaddress]{revtex4-2}

\usepackage{amsmath}
\usepackage{graphicx}
\usepackage[colorlinks=true, allcolors=blue]{hyperref}
\usepackage{ae}
\usepackage[T1]{fontenc}
\usepackage[ansinew]{inputenc}
\usepackage{amsmath}
\usepackage{amssymb}
\usepackage{graphicx}
\usepackage{color}
\begin{document}

\date{\today}

\title{Persistent current-carrying state of charge quasuparticles in $np$-ribbon featuring single Dirac cone}

\author{Anatoly M. Kadigrobov and Ilya M. Eremin }
\affiliation{Institut f\"ur Theoretische Physik III, Ruhr-Universitaet Bochum, D-44801 Bochum, Germany}


\begin{abstract}

The formation of persistent charge currents in mesoscopic systems remains an interesting and actual topic of condensed matter research. Here,  we analyze the formation of spontaneous arising persistent currents of charged  fermions in 2-dimensional 
electron-hole ribbons on the top and bottom of a 3-dimensional topological insulator. In such a device the two-dimensional Dirac fermions with  opposite chiralities are spatially separated that allows  these  currents to flow in  the opposite directions without compensating each other.  The nature of this phenomenon is based  on the  interference of the quasiparticle quantum waves which are  scattered with asymmetric  scattering phases  at the  lateral n-p chiral junction  and then reflected back  by the external boundaries of the  ribbon. As a result 
quasiparticles in the ribbon are shown to be in  unified electron-hole quantum 
states carrying the persistent current. 
\end{abstract}

\maketitle



Formation of the persistent currents in solid state systems have been always a subject of intense debates. Initially Felix Bloch demonstrated the impossibility of persistent electric currents in the ground many-body state of
interacting non-relativistic systems, known currently as a Bloch theorem\cite{Blochtheorem,Schmalian2010}. This theorem has been further extended to various cases of non-relativistic systems.\cite{Ohashi96,Zhang2019,Watanabe2019,Bachmann2021}
Interestingly enough, the idea of spontaneous currents was also actively discussed in the context of the chiral magnetic\cite{Vilenkin1980,Nielsen1983,Alekseev98,Fukushima08} and vortical\cite{Vilenkin79,Son2009,Landsteiner2011} effects. This yields an extension of the Bloch theorem to the relativistic systems, where it was shown that the chiral magnetic effect can be understood as a generalization of the Bloch theorem to a non-equilibrium steady state and the persistent axial currents are not prohibited\cite{Yamamoto2015}. 

At the same time, persistent current state with zero resistance were theoretically predicted to exist in the system having additional conserved charge.\cite{Else2021,Watanabe2022} Furthermore for the mesoscopic systems the persistent currents  are not prohibited by the Bloch theorem and is of continuous interest for the community.\cite{Huang2012,Sticlet2013,Tada2015,Zha2016,Miyawaki2018,Braak2019,Kobayashi2022,Chetcuti22}

\begin{figure}
\centerline{\includegraphics[width=\columnwidth]{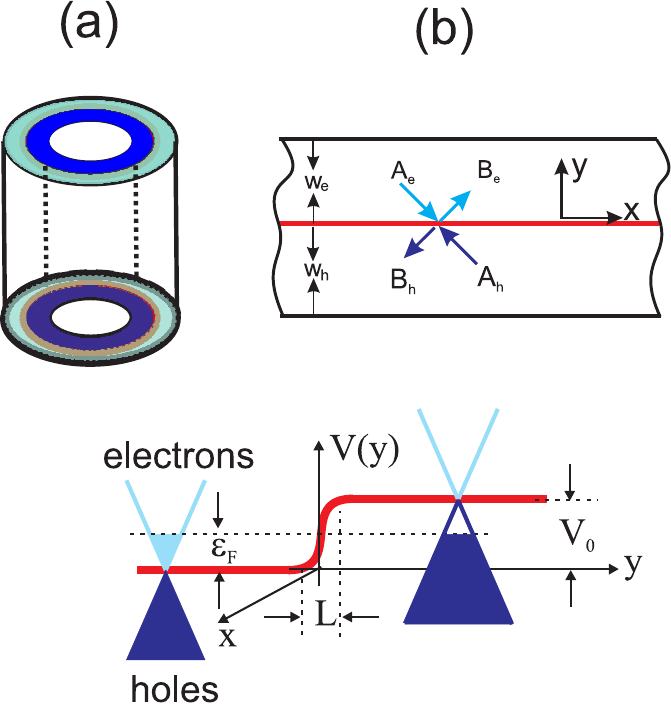}}
\caption{Schematic presentation of the proposed $np$-junction. The panel (a)  shows the proposed experimental geometry, consisting of two hollow cylinders (tubes) of finite length embedded into each other, made of a  topological insulator. (b) The thicknesses of the tubes are  $W_h$ and $W_e$ , respectively, and the surface Dirac fermions form on the top and the bottom rings of the tube. The length of the tube is chosen such that one could ignore the interaction of Dirac surface bands from upper and the lower rings of the tube. Due to applied gate voltage the surface Dirac fermions on the upper surface of the inner and outer cylinder are differently doped, forming $np$ junction.  (b) also  shows  the 2-channel scattering of quasiparticles at the $np$-junction on the top surface: incident electron and hole waves
with  amplitudes $A_e$ and $A_h$, respectively, tunnel through and reflected back with amplitudes $B_e$ and $B_h$ (the tunneling
is the interband one that transforms the electron into a hole and visa versa)}
\label{GateVolt}
\end{figure}

In this manuscript we analyze the formation of the persistent current carrying state in $n-p$-heterostructure made out of topological insulators, featuring odd number of two-dimensional Dirac fermions (schematically shown in Fig.\ref{GateVolt}).  Such mesoscopic devices with spatially separated Dirac fermion dispersions have been recently fabricated by various groups\cite{Kong2010,Cha2010,Zou2014,Jauregui2015,Parra2017,Liu2019}. Here, we show that such a mesoscopic structure can host the persistent charge current carrying state. 
In particular, the device we anticipate consists of two tubes of three-dimensional topological insulators with an electron-hole asymmetry of the surface Dirac fermions on the  surface  due to applied gate voltage. Then the two rings on the upper surface form electron-hole junction (np-junction), which  is schematically shown in Fig.\ref{GateVolt}). The  thincknesses   $W_e$ and $W_h$  of the electron and hole rings, respectively,
are assumed to satisfy the inequalities $W_e \sim W_h \gg \lambda_F $ where $\lambda_F = \hbar v/\varepsilon_F$ is the quasiparticle Fermi wave length
while $\varepsilon_F$ and $v$ are their Fermi energy and the corresponding velocity. An important feature is that the Dirac cone in each side of the $np$-junction is non-degenerate, which makes the topological insulator surface to be an ideal candidate for this junction.


Quantum dynamics of quasi-particles (electrons and holes) on the surface of TI with a lateral electron-hole junction (p-n junction)
is described by the 2-component envelope wave function $\check \psi$ satisfying the Dirac
equation
%
\begin{eqnarray}
\Big(  V(y)-\varepsilon \big)\psi_1 + v\big(p_x-\hbar\frac{d}{d y}\Big)\psi_2=0;\nonumber \\
 v\Big(p_x+ \hbar\frac{d}{d y}\big)\psi_1 +\big(  V(y)-\varepsilon \Big)\psi_2=0;
  \label{Schroedinger}
\end{eqnarray}
%
where the $V(y)$ is the lateral gate voltage
potential   extended along the
$x$-direction. Here, the axis $x$ is parallel to the sample external boundaries and the $np$-junction while the
$y$-axis is perpendicular to those as is shown in Fig.\ref{GateVolt}. 

Observe that Eq.(\ref{Schroedinger}) is itself symmetric with respect to the substitution of $p_x \to -p_x$. Indeed, there exists a unitary transformation matrix $U = \sigma_y, U^{\dagger} = \sigma_y$ that transforms $H(p_x,e)$  into $H(-p_x,e)$, which would imply $E(-p_x) = E(p_x)$. In what follows we show that the boundary conditions at the external ribbon  edges  as well as the condition of the electron and hole wave functions coupling at the np-junction change the situation.

Note that the boundary conditions at external sharp lattice edges for the above equation  were considered in various
papers (see, e.g., review paper \cite{Beenakker2008} and references there).
 Here we use those, derived in Ref. \cite{Kadigrobov2018}, with the usage of  the  $k \cdot p$ approximation \cite{Luttinger55}
 and analytical properties of Green functions of the Schr\"odinger equations  (this derivation is shortly presented in the Appendix A).

In particular,  the Dirac wave function is shown\cite{Kadigrobov2018} to be a superposition of the virtual Bloch states belonging to different momenta
and bands at the distances $l$ from the sample sharp edge which are less or of the same order as the atomic spacing $a$, $l  \lesssim  a$,
while at $l \gg a$ all the virtual states exponentially drop out from the superposition and the wave function reduces to the difference between
the incident  and the reflected Kohn-Lattinger Bloch functions \cite{Luttinger55} of the topological insulator surface sheet and hence  the boundary conditions for the Dirac equation
in the vicinity of  the ribbon edges $a \ll W_{e,h} - |y| $ can be written as follows:
%
\begin{eqnarray}
\check \psi(p_x, y) =  \hspace{4.2cm} \nonumber \\
C \left[\check \psi^{(in)}(p_x, y) - \check \psi^{(out)}(p_x, y)  \right]
 \label{boundcond}
\end{eqnarray}
%
where $C$ is a constant and $\check \psi^{(in, out)}(p_x, y)$  are solutions of Eq.(\ref{Schroedinger}) for quasiparticle waves
incoming  and outgoing from
the external boundaries (see Appendix A).

In order to find the exact form of the boundary condition at the np-junction  we assume that the characteristic change interval $L$ of  the gate voltage $V(y)$ satisfies the condition $a \ll L \ll \lambda_F$
that allows to  use Eq.(\ref{Schroedinger}) in which $V(y)$ is approximated
by  the step-function of the height $V_0$ (see Fig.\ref{GateVolt}):
\begin{eqnarray}
V(y)= \Big\{\begin{array}{lc} 0 & y\leq 0 \\
V_0 & y \geq 0;
\end{array}
\end{eqnarray}
%
The general form of the solution of Eq.(1)  can be expressed for holes ($y \geq 0$)
\begin{eqnarray}
\check \Psi_h=A_h\frac{\sqrt{V_0-\varepsilon}}{v \sqrt{2 p_h}}\left(\begin{array}{c}
 1 \\
-\displaystyle \frac{v (p_x+i p_h)}{V_0-\varepsilon}
\end{array}\right)e^{iy p_h } \nonumber \\
+B_h\frac{\sqrt{V_0-\varepsilon}}{v \sqrt{2 p_h}}\left(\begin{array}{c}
 1 \\
-\displaystyle \frac{v (p_x-i p_h)}{V_0-\varepsilon}
\end{array}\right)e^{-i y p_h }
\label{Psi1A}
\end{eqnarray}
and the electrons ($y \leq 0$)
\begin{eqnarray}
\check \Psi_e=A_e\frac{\sqrt{\varepsilon}}{v \sqrt{2 p_e}}\left(
\begin{array}{c}
 1 \\
\displaystyle \frac{v (p_x+i p_e)}{\varepsilon}
\end{array}\right) e^{i y p_e } \nonumber \\
+B_e\frac{\sqrt{\varepsilon}}{v \sqrt{2 p_e}}\left(\begin{array}{c}
 1 \\
\displaystyle \frac{v (p_x-i p_e)}{\varepsilon}
\end{array}\right)e^{-i y p_e }
\label{Psi2A}
\end{eqnarray}
where $A_{e,h}$ and $B_{e,h}$ are the constants and
%
\begin{eqnarray}
p_e=\sqrt{\left(\frac{\varepsilon}{v}\right)^2-p_x^2}; \;\; p_h=\sqrt{\left(\frac{V_0-\varepsilon}{v}\right)^2-p_x^2}
 \label{p_e,h}
\end{eqnarray}
%
To find the constants we match the electron and the hole wave functions at the interface of the $np$-junction $y=0$ and take into account the signs of the electron $v_y^{(e)} = v^2p_y/\varepsilon$ and the hole velocity
$v_y^{(h)}=-v^2p_y/(V_0-\varepsilon)$, respectively. The set of algebraic 
equations to determine the constants at the electron and hole wave function  inside the region  $\varepsilon \leq V_0/2$, $|p_x| \leq \varepsilon/v$ and $V_0/2 \leq  \varepsilon \leq V_0/$, $|p_x| \leq (V_0 - \varepsilon)/v$ have the form:
\begin{eqnarray}
\bar{A}_h + \bar{B}_h = \bar{A}_e + \bar{B}_e;     \hspace{0.7cm} \nonumber \\
 \frac{p_x+i p_h}{\varepsilon -V_0} \bar{A}_h+\frac{p_x-i p_h}{\varepsilon} \bar{B}_h \nonumber \\
=\frac{p_x+i p_h}{V_0-\varepsilon} \bar{A}_e+ \frac{p_x-i p_h}{V_0-\varepsilon} \bar{B}_e;
\label{algebraicEq}
\end{eqnarray}
where
\begin{eqnarray}
\bar{A}_h=\sqrt{\frac{V_0-\varepsilon}{2 p_h}}A_h; \;\; \bar{B}_h=\sqrt{\frac{V_0-\varepsilon}{2 p_h}}B_h;\nonumber \\
\bar{A}_e=\sqrt{\frac{\varepsilon}{2 p_e}}A_h; \;\; \bar{B}_e=\sqrt{\frac{\varepsilon}{2 p_e}}B_e;\hspace{1.1cm}
\label{ABbarA}
\end{eqnarray}

Outside of the above-mentioned area, $\varepsilon \leq V_0/2,  \; (V_0-\varepsilon)/v \geq |p_x| \geq \varepsilon/v,$ and  $\varepsilon \geq (V_0/2,  \; (V_0-\varepsilon)/v \leq |p_x| \leq \varepsilon/v$, holes 
undergo the complete internal reflections at the np-junction.
For the hole momentum and the energy  inside the indicated limits  one finds that the electron momentum projection $p_y$ is imaginary, $p_y^{(e)}=i \sqrt{p_x^2-(\varepsilon/v)^2}$, and hence the electron wave function (into which the incident hole is transformed after passing through
 the np-junction) exponentially  decays inside the electronic part of the ribbon, $y <0$.
Taking this into account we find at the interface of the $np$-junction, $y=0$:
\begin{eqnarray}
\bar{A}_h + \bar{B}_h ={B}_e \hspace{2.1cm} \nonumber \\
-\bar{A}_h \frac{p_x +i p_h}{V_0-\varepsilon}  -\bar{B}_h \frac{p_x -i p_h}{V_0-\varepsilon} = \bar{B}_e \frac{p_x -i p_e}{\varepsilon}
\label{ABhBeA1}
\end{eqnarray}
%
Next, solving equations Eq.(\ref{algebraicEq})  one finds the $2 \times 2$ unitary matrix  $\hat{\rho}$  that connect the constant factors $A_e,\;A_h$ at
the incident and $B_e \; B_h$ at the outgoing wave
functions (see also Fig.1(b)):
%
\begin{eqnarray}
\hat{\rho}= e^{i\varphi}\left(%
\begin{array}{cc}
 |r|e^{i \mu} & t\\
-t^{\star}  & |r|e^{i \mu} \\
\end{array}%
\right)
 \label{Matrix}
\end{eqnarray}
%
where
%
\begin{eqnarray}
|r| =\left\{\frac{(V_0 p_x)^2 +[p_e(V_0-\varepsilon )-p_h \varepsilon]^2}{(V_0 p_x)^2 +[p_e(V_0+\varepsilon )+p_h \varepsilon]^2} \right \}^{1/2}
 \label{r}
\end{eqnarray}
%
and
%
\begin{eqnarray}
t = \frac{2 \sqrt{p_e p_h \varepsilon (V_0 -\varepsilon)}}{\sqrt{(V_0 p_x)^2 + [p_e (V_0 - \varepsilon) + p_h \varepsilon]^2}}
 \label{t}
\end{eqnarray}
%
are the probability amplitudes for an incident quasi-particle to be reflected back or to be transmitted
through the barrier, respectively, while
\begin{eqnarray}
\varphi =\pi+ \arctan \frac{ p_e  (V_0 -\varepsilon) +p_h \varepsilon }{V_0 p_x}
 \label{phi}
\end{eqnarray}
%
 and
\begin{eqnarray}
\mu = \arctan \frac{ p_e  (V_0 -\varepsilon) - p_h \varepsilon }{V_0 p_x}
 \label{mu}
\end{eqnarray}
are the scattering phases. Most importantly, as follows from Eqs.(\ref{phi})-(\ref{mu}) that scattering phases are asymmetric, i.e. $\varphi(p_x)\neq \varphi(-p_x), \;\;\mu(p_x)\neq \mu(-p_x)$. This will be the origin current carrying state.

Note, the unitary scattering matrix $\hat{\rho}$ describes inter-band electron-hole tunneling that takes place inside
the area defined by the following inequalities: $\varepsilon \leq V_0/2$, $|p_x| \leq \varepsilon/v$ and $V_0/2 \leq  \varepsilon \leq V_0/$, $|p_x| \leq (V_0 - \varepsilon)/v$  which is seen as a diamond in Fig.2.  The boundaries of this area are defined by the condition that the transverse momenta $p_e$ and $p_h$ are real.

Matching the solutions of the Dirac equation, Eqs.(\ref{Psi1A})-(\ref{Psi2A}),  with the usage of the boundary condition at the sample edges, Eq.(\ref{boundcond}),  and the scattering matrix at the np-junction, Eq.(\ref{Matrix}), one finds
the dispersion equation for the quasi-particles in the n-p junction as follows. (1) Inside the energy and  momentum intervals $\varepsilon \leq V_0/2, \;\; |p_x| \leq \varepsilon/v$ and $\varepsilon \geq (V_0-\varepsilon)/2, \;\; |p_x| \leq (V_0-\varepsilon)/v$ the dispersion equation is
%
\begin{eqnarray}
D(\varepsilon, p_x) \equiv \hspace{5.2cm}\nonumber \\
\cos\left[\Phi_e -\Phi_h+\varphi\right]+
|r|\cos\left[\Phi_e+\Phi_h+\mu\right]=0;
\label{D}
\end{eqnarray}
%
where
%
\begin{eqnarray}
 \Phi_e (\varepsilon, p_x) =\frac{W_e p_e(\varepsilon, p_x)}{\hbar}; \;\;  \Phi_h (\varepsilon, p_x) =\frac{W_h p_h(\varepsilon, p_x)}{\hbar}.
 \label{p}
\end{eqnarray}
(2) Outside the above-mentioned intervals the quasi-particles undergo the complete internal reflections  at the np-junction, 
Eq.(\ref{ABhBeA1}), and  their dispersion equations are 
\begin{eqnarray}
D_h(\varepsilon, p_x)=\sin[\Phi_h-\varphi_h]; \;\;|p_x| \geq \frac{\varepsilon}{v},
 \; \varepsilon \leq\frac{V_0}{2}  \hspace{0.7cm} \nonumber \\
D_e(\varepsilon, p_x)=\sin[\Phi_e+\varphi_e]; \;\;|p_x| \geq \frac{V_0-\varepsilon}{v},
 \; \varepsilon \geq\frac{V_0}{2}
\label{Deh}
\end{eqnarray}
where
\begin{eqnarray}
\varphi_h&=& \pi+\arctan\left[\frac{p_h \varepsilon }{V_0p_x+ |p_e|(V_0-\varepsilon)}\right]\nonumber \\
\varphi_e&=& \pi+ \arctan\left[\frac{ p_e(V_0-\varepsilon)}{(V_0 p_x-|p_h|\varepsilon)}\right]
 \label{phi_eh}
\end{eqnarray}

\begin{figure}
\centerline{\includegraphics[width=4cm]{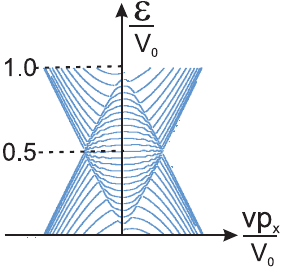}}
\caption{Calculated spectrum of quasiparticles in the np-ribbon, $\varepsilon_n(p_x)$. Numerical calculations are performed
for the parameter $\Lambda = 50$. Inside the diamond ($\varepsilon \leq 0.5 V_0$, $|p_x| \leq \varepsilon /v$ and
$\varepsilon \geq 0.5 V_0$, $|p_x| \leq (V_0 -\varepsilon )/v$) the spectrum corresponds to
the unified electron-hole quantum interfering state, the energy bands being asymmetric, $\varepsilon_n(p_x) \neq \varepsilon_n(-p_x)$,
  and rather flat.
Outside the diamond the spectrum is of the holes, $\varepsilon <0.5 V_0$, and electrons, $\varepsilon > 0.5 V_0$
because the corresponding quasiparticles undergo total reflections at the np-junction.}
\label{spectrum}
\end{figure}

Solutions of Eq.(\ref{D})-( \ref{Deh}) give the dispersion law of the electron-hole quasiparticles in
the n-p ribbon, $\varepsilon_n (p_x)$,  presented in
Fig.\ref{spectrum}
As follows from Eqs.(\ref{D}),(\ref{Deh}) quasiparticles in the np-ribbon are electron-hole coupled states only inside the diamond
($\varepsilon \leq 0.5 V_0$, $|p_x| \leq \varepsilon /v$ and
$\varepsilon \geq 0.5 V_0$, $|p_x| \leq (V_0 -\varepsilon )/v$) shown
in Fig.\ref{spectrum}. Outside the diamond, $\varepsilon \leq 0.5 V_0$, $|p_x| \geq \varepsilon /v$ and
$\varepsilon \geq 0.5 V_0$, $|p_x| \geq (V_0 -\varepsilon )/v$, electrons and holes undergo the complete
internal reflections at the np-junction and their dispersion equations are defined by Eq.(\ref{Deh}). Inside the diamond, the spectrum   consists of a series of  extremely narrow bands of
the width $\Delta \varepsilon \sim 2 \pi \hbar v/W$ and the dispersion depends on the variation of
the  momentum, $\Delta p_x \sim \hbar/W$.
Such a peculiar spectrum is the result of the quantum interference of the electron and hole waves.

As this quantum interference is associated
with  the 2-channel scattering of the waves  at the n-p junction (which breaks the symmetry $p_x \rightarrow -p_x$
of the scattering  phases, see  Eq.(\ref{D}) and Eqs.(\ref{phi},\ref{mu})) it
results in the asymmetry of the new quasiparticle velocity
 $\bar{v}_x(\varepsilon, p_x) \neq \bar{v}_x(\varepsilon, -p_x)$ where
%
\begin{eqnarray}
\bar{v}_x =\frac{d \varepsilon_n(p_x)}{d p_x}= - \frac{\partial D/\partial p_x}{\partial D/\partial \varepsilon},
\;\;\;D(\varepsilon, p_x)=0
 \label{velocity}
\end{eqnarray}
%
As one easily sees from Eq.(\ref{D}, \ref{phi},\ref{mu})  the dispersion equation  $D(\varepsilon, p_x)=0$  is asymmetric with respect to the momentum inversion, $p_x \rightarrow -p_x$   and hence the spectrum of electron-hole quasi-particles 
is asymmetric as well, $\varepsilon_n(p_x) \neq \varepsilon_n(-p_x)$. Therefore, velocities $v_x(p_x) \neq v_x(-p_x)$ and quasiparticles with opposite directions of  $p_x$-momenta  carry  the probability
 density currents of  unequal magnitudes. As a result, the quantum ground state of the electron-hole Fermi gas
 in the $np$ ribbon or a ring  carries a persistent current  in the absence of any external fields, which is rather remarkable result.

 It is worth to note that the phenomenon under consideration is absent in an electron or hole graphene ribbon in which the
 electron-hole interference is absent. In this case the corresponding  dispersion laws are  symmetric  and the probability currents  flowing in the opposite directions  compensate each other.
Note,  in contrast to the
conventional persistent current under magnetic field the predicted  phenomenon
 is based on the combination of the quantum
interference  which arises after 2-channell scattering of the quasi-particle waves at the np-junction and then reflected back
by the external ribbon boundaries, and the asymmetry of the scattering phases
 with respect to $p_x \rightarrow -p_x$ , see Eqs.(\ref{phi})-( \ref{mu}). This also puts a constraint on the observability of the effect, which requires a mesoscopic size of the junction and its good transparency.

Expression for the peculiar persistent current under consideration can be obtained in the following form:
\begin{eqnarray}
j_{ps}=-2 e Tr[\hat{v}_xf_0(\hat{H})]\nonumber \\
=- \frac{ e  }{ \pi \hbar}
\int d \varepsilon f_0\left[\varepsilon\right]\int d p_x
\bar{v}_x\left[\varepsilon, p_x\right]  \sum_n   \delta\left[ \varepsilon-\varepsilon_n(p_x)\right]
 \label{current1}
\end{eqnarray}
%

 To calculate it we use the approach developed previously for magnetic breakdown systems \cite{Slutskin1970}. Namely, using the  identity
 %
\begin{eqnarray}
\sum_n   \delta\left[\varepsilon -\varepsilon_n(p_x)\right]=\left| \frac{\partial D(\varepsilon, p_x)}{\partial \varepsilon}\right|\delta[ D(\varepsilon, p_x)]
 \label{identity}
\end{eqnarray}
together with Eq.(\ref{velocity}) one presents   Eq.(\ref{current1}) in the following form:
\begin{eqnarray}
j_{ps}=
- \frac{e }{ \pi \hbar}
\int d \varepsilon f_0\left[\varepsilon\right]\int d p_x
\frac{\partial D(\varepsilon, p_x)}{\partial p_x}\delta\left[ D(\varepsilon, p_x)\right]
 \label{current2}
\end{eqnarray}

As one sees from Eq.(\ref{p_e,h})  $p_e(\varepsilon) \leq p_h(\varepsilon)$ if $\varepsilon \leq  V_0/2$
and $p_h(\varepsilon) \leq p_e(\varepsilon)$ if $\varepsilon \geq  V_0/2$. On the other hand $D(\varepsilon, p_x )$
in the integrand of Eq.(\ref{current2}) has a jump  $p_x=\pm 0$. Taking it into account one re-writes Eq(\ref{current2}) as follows:
\begin{eqnarray}
\frac{j_{ps}}{e/\pi \hbar}=
- \int_0^{\infty} d \varepsilon f_0\left(\varepsilon\right)
\left\{\Theta \left[D(\varepsilon, -0)\right]- \Theta \left[D(\varepsilon, +0)\right]  \right\}\nonumber \\
 +\int^{V_0/2}_{0}d \varepsilon f_0\left(\varepsilon\right)\Big\{ \int_{-p_h^{(m)}}^{-p_e^{(m)}}
 \frac{d \Theta[D_h]}{d p_x} d p_x \nonumber \\
 +\int_{p_e^{(m)}}^{p_h^{(m)}}
 \frac{d \Theta[D_h]}{d p_x} d p_x \Big \}
  + \int^{V_0}_{V_0/2}d \varepsilon f_0\left(\varepsilon\right)\left\{... \right\}\hspace{2.2cm}
 \label{current3}
\end{eqnarray}
where
\begin{eqnarray}
p_e^{(m)}=\frac{\varepsilon}{v}; \;\;\;
p_h^{(m)}=\frac{V_0-\varepsilon}{v}
\label{maxmomentums}
\end{eqnarray}
are the maximal electron and hole momenta, respectively while
according to 
Eq.(\ref{phi}-\ref{D}) one has
\begin{eqnarray}
D(\varepsilon, \pm 0)=\cos\left[\frac{W(V_0 - 2 \varepsilon )}{\hbar v} \mp \frac{\pi}{2}\right]
\label{Dappendix}
\end{eqnarray}

Using the equalities   $D(\varepsilon, p_e^{(m)})=D_h(\varepsilon, p_e^{(m)}) $  and    $D_h(\varepsilon, -p_h^{m}) =D_h(\varepsilon, p_h^{m})$   valid at $\varepsilon \leq V_0/2$, and $D(\varepsilon, p_h^{(m)})=D_h(\varepsilon, p_h^{(m)}) $  and    $D_e(\varepsilon, -p_e^{m}) =D_e(\varepsilon, p_h^{e})$   valid at $\varepsilon \geq V_0/2$ (see Eqs.(\ref{D}, \ref{Deh}) one finds the final expression for
the quantum interference persistent current flowing in the absence of external fields  as follows:
\begin{eqnarray}
\frac{j_{ps}}{j_0}=
\int_{0}^{1}f_0(z) \Big\{\Theta\Big[\sin \big(\Lambda(1-2z)\big)\Big] \hspace{2.5cm}\nonumber \\
-\Theta\big[-\sin \big(\Lambda(1-2z)\big)\big] \Big\} d z\nonumber \\
 \label{CurrentFinal}
\end{eqnarray}
where $\Theta (z)$ is the step function and
\begin{eqnarray}
f_0(z)= \Big(\exp\frac{(z- \varepsilon_F/V_0)}{T/V_0}+1\Big)^{-1}\nonumber \\
j_0=\frac{e V_0}{\pi \hbar}; \;\;\Lambda =\frac{WV_0}{\hbar v} \hspace{2cm}
 \label{parameters}
\end{eqnarray}
Here $f_0$  is the normalized Fermi distribution function, $j_0$ is the density of the current carried by the quasiparticles  in the junction which has $p_x$-momentum of the same sign; $\Lambda \gg 1$ is  the semiclassical parameter. For the sake of simplicity, in what follows we consider
$W_e=W_h=W$.
%
\begin{figure}
\centerline{\includegraphics[width=8cm]{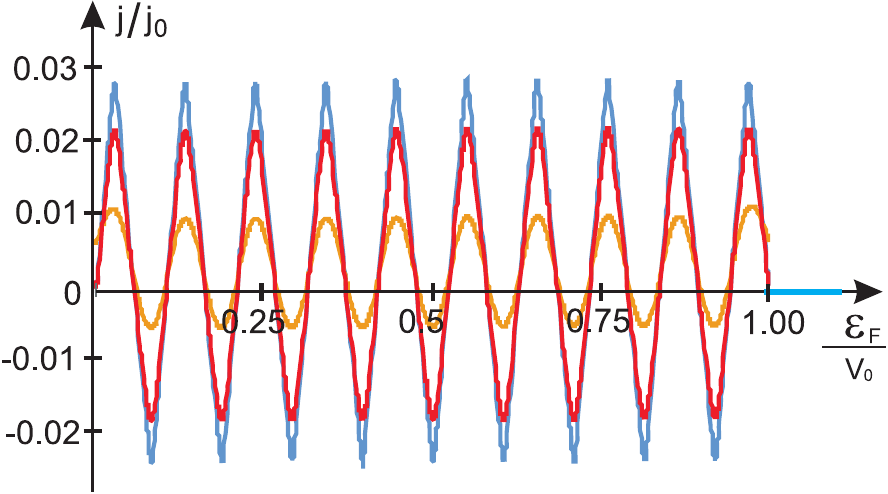}}
\caption{Calculated dependence on
$\varepsilon_F/V_0$  of the  current $j_{ps}$ flowing along the np-ribbon in the absence of external fields. Numerical calculations are performed  for  $\Lambda= W V_0/ \hbar v = 30$. Curves of the largest and smallest
 amplitudes (blue and orange on-line) are  at $T=0$  and  $T=10^{-2}V_0$, respectively.  The curve of the middle amplitude (red on-line) is
 for a dirty graphene $W_e+W_h=12 l_0$ at   $T=0$   ($l_0$ is the quasiparticle free path length).
The normalization parameter $j_0 = e V_0/\pi \hbar$ is of the order of the value of the current carried by the quasiparticles
of the ribbon moving in the same direction.}
\label{current}
\end{figure}
%
The typical dependence of the current on the ratio between the Fermi energy or  gate voltage, $\varepsilon_F/V_0$, is
presented in Fig.\ref{current}.

The  integrand  in the right-hand side of Eq.(\ref{CurrentFinal}) comes from the jump of  $D(\varepsilon, p_x)$  at $p_x =\pm 0$
(see Eqs.(\ref{phi},\ref{mu},\ref{D})).
The dependence of $j_{ps}$ on $\varepsilon_F/V_0$   displays  an oscillating character both in the current direction and the amplitude, the latter
being proportional to $j_0/\Lambda$.   This decrease of the current amplitude by the semiclassical parameter $\Lambda \gg 1$
is deterimed by the fast semiclassical
oscillations of the integrand in Eq.(\ref{CurrentFinal}).

As the phenomenon under consideration is  based on the asymmetry of  the phases of the scattering amplitudes  at the np-junction
this persistent current has weak dependence  on the scattering by impurities
provided  the latter (being symmetric) does not change the problem
symmetry and hence the current. To show this we
take into account the impurities by the simple broadening  the $\delta$- and $\Theta$-functions in Eq.(\ref{current1},\ref{current2}, \ref{CurrentFinal})
by $\gamma = \hbar v/l_0$  (the electron-impurity free path length is $l_0$) and numerically calculated modified Eq.(\ref{CurrentFinal}) for the ribbon width
 $W_e+W_h=12 l_0$   (see Fig.\ref{current}).
In perfect analogy to the  conventional persistent current under magnetic field\cite{Imry}
the predicted current survives at distances $\lesssim l_\varphi$    at which  the quasiparticle wave function phase conserves.

The spontaneous current under consideration can flow along a closed  np-ribbon in the form of a ring formed on the top of the topological insulators  cylinders
the radius of which satisfies the inequality  $W \ll R \lesssim l_{\varphi}$) (see Fig.\ref{GateVolt}a,b).
In a  np-ribbon of the length $l_{rib} \lesssim l_{\varphi}$ the spontaneous current is absent but
there is a voltage drop  between the ribbon ends,  $V_r =j_{ps} l_{rib}/\sigma$
where  $\sigma$ is the conductivity of the ribbon. As follows from the above  analysis irradiation of the surface of the topological insulator top surface ring 
np-ribbin excites the additional current  predominantly flowing in one direction that may be used for generation of a substantial
 photocurrent and the photovoltaic effect in single cone Dirac systems, (for  investigations of the problems  see Refs. \cite{ChanPhoto2017,Mclver2012,Kuroda2016}).

%
%

To conclude, we analyze  a peculiar persistent  current formation along a mesoscopic electron-hole ribbon, formed on the surface of the topological insulator.  Such currents may form even in the absence of both the magnetic and electric fields and arise due to odd number of Dirac cones, electron-hole asymmetry of the junction and finite size effects. 
We expect that in the case of the topological insulator (shown in Fig.\ref{GateVolt},a) the Dirac fermions 
at the top and  bottom of the sample have the opposite chiralities and hence their persistent 
charge currents flow in the opposite directions.  
Being spatially separated these planar currents form the anti-Helmholtz coil (widely used for creating the cusp traps for ultra cold atoms) producing  
axially symmetric  magnetic field in the form of the cusp around the middle of the  coil axis.  
Therefore, the presence of the spontaneous persistent charge currents in such a topological insulator may be 
detected by measuring this specific  magnetic field.

\begin{appendix}

\section{Dirac quasiparticle dynamics and boundary conditions}

In this section dynamics of quasi-particles and boundary conditions  at the sharp lattice edge based on the k.p-approximation is  shortly
described (their detailed analysis is presented in Ref.\cite{Kadigrobov2018}).

Derivation of the boundary conditions is performed in two steps: 1)using the k.p-approximation\cite{Luttinger55}  (the Lattinger-Kohn model)
one derives the Dirac equation for the envelope wave functions; 2) using the Green functions for the Schr\"odinger equation expanded
in terms of the Bloch functions of all bands and using the found connection between the Bloch functions and the  envelope
functions one derives the boundary conditions sought for the envelope functions near the lattice sharp edge.

{\bf Step 1}. The Schr\"odinger equation for  noninteracting  quasiparticles   is written as
\begin{eqnarray}
\left[-\frac{\hbar^2}{2 m}\frac{\partial^2}{\partial\mathbf{ r}^2} +U(\mathbf{r})\right] \varphi_{s,\mathbf{p}}(\mathbf{r})=\varepsilon_s(\mathbf{p})
\varphi_{s,\mathbf{p}}(\mathbf{r})
 \label{Schroedinger0A}
\end{eqnarray}
where $U(\mathbf{r})=U(\mathbf{r}+\mathbf{a})$ is the lattice periodic  potential ($\mathbf{a}$ is
the lattice vector)
and
\begin{eqnarray}
 \varphi_{s,\mathbf{p}}(\mathbf{r})=e^{i\frac{\mathbf{pr}}{\hbar}}u_{s,\mathbf{p}}(\mathbf{r})
 \label{BlochA}
\end{eqnarray}
is the Bloch function while $u_{s,\mathbf{p}}(\mathbf{r})$ is its periodic factor, $\mathbf{p}$ is
the electron quasi-momentum while $\varepsilon_s(\mathbf{p})$ is the dispersion law and $s=1,2$ are the band numbers
of the  electron and hole bands degenerated at $\mathbf{p}=0$.


In the presence of a potential $V(\mathbf{r})$ the  Schr\"odinger equation reads
\begin{eqnarray}
\left[-\frac{\hbar^2}{2 m}\frac{\partial^2}{\partial\mathbf{ r}^2} +U(\mathbf{r}) +
V(\mathbf{r}) \right] \Psi(\mathbf{r})=\varepsilon
\Psi(\mathbf{r})
 \label{SchroedingerVA}
\end{eqnarray}
Here and below we asume that the potential $V(\mathbf{r})$ in a smooth function of the atomic scale.

To solve the problem we expand the wave function
in terms  of the modified Bloch functions (the Luttinger-Kohn  functions)
\begin{eqnarray}
\chi_{\alpha,\mathbf{p}}=\exp{\left(i\frac{\mathbf{pr}}{\hbar}\right)}
u_{s,0}(\mathbf{r})
 \label{chiA}
\end{eqnarray}
 as follows:
\begin{eqnarray}
\Psi(\mathbf{r})=\sum_{s=1,2} \int g_{s,\mathbf{p}}\chi_{s,\mathbf{p}}(\mathbf{r})d\mathbf{p}
 \label{expansionA}
\end{eqnarray}

Using  Eq.(\ref{expansionA}) and the above-mentioned degeneration perturbation theory ("k.p-method") one derives
the Dirac equation (see Ref.\cite{Kadomtsev1968} for details)
\begin{eqnarray}
\left(%
\begin{array}{cc}
  V(\mathbf{r}) &  v \left(p_x -\hbar \displaystyle \frac{ d}{d y} \right)\\
 v \left(p_x +\hbar \displaystyle \frac{ d}{dy} \right) &V(\mathbf{r})\\
\end{array}%
\right)\left(%
\begin{array}{c}
  \psi_1 \\
  \psi_2\\
\end{array}%
\right) \nonumber \\
=\varepsilon \left(%
\begin{array}{c}
  \psi_1 \\
  \psi_2\\
\end{array}%
\right)
 \label{DiracEquationA}
\end{eqnarray}
for the envelope functions
\begin{eqnarray}
\psi_{1,2}(\mathbf{r})=\int g_{1,2}(\mathbf{p})\exp\{i \frac{p \mathbf{r}}{\hbar}\}d
\mathbf{r}
 \label{DiracA}
\end{eqnarray}

As one sees from Eq.(\ref{DiracA}, \ref{chiA}) the wave functions Eq.(\ref{expansionA}) may be re-written in
terms of the envelope functions as follows:
\begin{eqnarray}
\Psi(\mathbf{r})=\sum_{s=1,2} u_{s,0}(\mathbf{r})\psi_s (\mathbf{r})
\label{EnvelopeWFA}
\end{eqnarray}

{\bf Step 2}. In order to find boundary conditions for the Dirac equation we investigate Schr\"odinger equation  Eq.(\ref{Schroedinger0A})
(in which $\varphi_{s,\mathbf{p}}(\mathbf{r})$ is changed to $\Psi(\mathbf{r})$) with the following boundary conditions:
\begin{eqnarray}
\Psi(\mathbf{r})&=&0, \hspace{1.4cm} y=0  \nonumber \\
\Psi (\mathbf{r}) &=& \varphi_{s,\mathbf{p}_0}^{(gr,in)}(\mathbf{r}) , \;\; y\rightarrow
+\infty
 \label{BoundaryCondition1A}
\end{eqnarray}
Here $\varphi_{s_0,\mathbf{p}_0}^{(gr,in)}$ is the  Bloch function   incident to the boundary; this function  is presented in Eq.(\ref{EnvelopeWFA})
in which the envelope functions $\psi_s (\mathbf{r})$ are solutions of Eq.(\ref{DiracEquationA}) at $V(\mathbf{r})=0$
and $\mathbf{p}_0=(p_x,p_y^{(in)})$ where
$p_x$   is  the conserving momentum projection on the edge and $p_y^{(in)}=-\sqrt{(\varepsilon/v)^2-p_x^2}$.

To solve  the  problem of reflection by the abrupt edge at $y=0$, Green's function
for  Schr\"odinger equation Eq.(\ref{Schroedinger0A}) is used:
\begin{eqnarray}
\left(-\frac{\hbar^2}{2 m}\frac{\partial^2}{\partial\mathbf{ r}^2} +U(\mathbf{r})
-\varepsilon\right)G(\mathbf{r},\mathbf{r}^{\prime})=\delta(\mathbf{r}-\mathbf{r}^{\prime})
 \label{GreenFunctionEqA}
\end{eqnarray}
in which the  lattice potential $U(\mathbf{r})$ covers the whole plane $(x,y)$.

 Expanding
$G(\mathbf{r},\mathbf{r}^{\prime})$  in the series of Bloch wave functions   and using Eq.(\ref{GreenFunctionEqA}) one  finds
\begin{eqnarray}
G(\mathbf{r},\mathbf{r}^{\prime})= \sum_{s=1,2}\int\frac{\varphi_{s,\mathbf{p}}^{(gr)\star}(\mathbf{r})
\varphi_{s,\mathbf{p}}^{(gr)}(\mathbf{r}^{\prime})}{\varepsilon-\varepsilon_s^{(gr)}(\mathbf{p})+i \delta}d
\mathbf{p} \nonumber \\
 +\sum_{s \neq 1,2}\int\frac{\varphi_{s,\mathbf{p}}^{\star}(\mathbf{r})
\varphi_{s,\mathbf{p}}(\mathbf{r}^{\prime})}{\varepsilon-\varepsilon_s(\mathbf{p})+i \delta}d
\mathbf{p}
 \label{GreenFunctionExpandedA}
\end{eqnarray}
where the summation goes over all energy bands  while $\varepsilon_s^{(gr)}(\mathbf{p})= \pm v p$ is the graphene dispersion and
 $\delta \rightarrow +0$


Taking into account analytical properties of the virtual and graphene energy bands as functions of complex momenta and using Eqs.(\ref{Schroedinger0A}),(\ref{BoundaryCondition1A}), and (\ref{GreenFunctionExpandedA}) together with Eq.(\ref{EnvelopeWFA})
we find that at $y \gg a$ (here $a$ is the characteristic  value of the period of the lattice) the   wave fuction
is the difference between the incident and outgoing Bloch functions of the infinite sample:
\begin{eqnarray}
\Psi_{p_x}(\mathbf{r})=C \left( \varphi_{s;p_x, p_y^{(in)}}^{(gr)}(\mathbf{r}) -  \varphi_{s;p_x, p_y{( out)}}^{(gr)}(\mathbf{r})\right)
 \label{GrapeneWVBoundaryA}
\end{eqnarray}
where $p_y^{(in)}$ and $p_y^{(out)} = -p_y^{(in)}$ are the $y$-projections of the quasiparticle momentum while $C$ is the normalizing
constant (details of calculations are given in Ref. \cite{Kadigrobov2018}).

From Eq.(\ref{GrapeneWVBoundaryA}) and Eq.(\ref{EnvelopeWFA}) one easily finds that at the distanses from the lattice sharp edge much larger than the atomic spacing, $l\gg a$,  the envelope function $\check{\Psi}(\mathbf{r})$ is the difference between  the incident and outgoing wave functions (which are
 two independent solutions of the Dirac equation
Eq.(\ref{DiracEquationA})):
\begin{eqnarray}
\check{\Psi}(\mathbf{r})= e^{ix p_x}\left[e^{iy p_y^{(in)}}\left(%
\begin{array}{c}
  1\\
 e^{i \varphi}\\
\end{array}%
\right)-  e^{-i(y p_y^{(in)})} \left(%
\begin{array}{c}
1 \\
   e^{-i \varphi}\\
\end{array}%
\right)  \right] \nonumber\\
 \label{GraphBCA}
\end{eqnarray}
where the phase $\varphi =\arctan (p_y^{(in)}/p_x)$.

\end{appendix}

\bibliography{references}

\end{document}